%% file: HTGWv2.0.tex
\documentclass[reprint,showpacs,showkeys,superscriptaddress,prb]{revtex4-1}

\usepackage{amssymb}
\usepackage{amsmath}
\usepackage{graphicx}
\usepackage{booktabs}
\usepackage{longtable}

\DeclareGraphicsExtensions{.eps}
\newcommand{\GG}{{\bf G}}
\newcommand{\kk}{{\bf k}}

\newcommand{\rr}{{\bf r}}
\newcommand{\occ}{{\text{occ}}}
\newcommand{\qq}{{\bf q}}
\newcommand{\qG}{{\bf q+G}}
\providecommand{\abs}[1]{{\lvert#1\rvert}}

\newcommand{\BZ}{{{\text{BZ}}}}
\input{newcommands.tex}
\AtBeginDocument{
\heavyrulewidth=.08em
\lightrulewidth=.05em
\cmidrulewidth=.03em
\belowrulesep=.65ex
\belowbottomsep=0pt
\aboverulesep=.4ex
\abovetopsep=0pt
\cmidrulesep=\doublerulesep
\cmidrulekern=.5em
\defaultaddspace=.5em
}

\begin{document}

%\begin{frontmatter}

\title{Automation methodologies and large-scale validation for $GW$, towards high-throughput $GW$ calculations.}

%High-throughtput $GW$: automatization methodologies and large-scale validation}

%\address[ucl]{Nanoscopic Physics, Institute of Condensed Matter and Nanosciences, Universit\'{e} Catholique de Louvain, 1348 Louvain-la-Neuve, Belgium}

\author{M. J. {van Setten}}
\email{mjvansetten@gmail.com}  
\affiliation{Nanoscopic Physics, Institute of Condensed Matter and Nanosciences, Universit\'{e} Catholique de Louvain, 1348 Louvain-la-Neuve, Belgium}
\author{M. Giantomassi}
\affiliation{Nanoscopic Physics, Institute of Condensed Matter and Nanosciences, Universit\'{e} Catholique de Louvain, 1348 Louvain-la-Neuve, Belgium}
\author{X. Gonze}
\affiliation{Nanoscopic Physics, Institute of Condensed Matter and Nanosciences, Universit\'{e} Catholique de Louvain, 1348 Louvain-la-Neuve, Belgium}
\author{G.-M. Rignanese}
\affiliation{Nanoscopic Physics, Institute of Condensed Matter and Nanosciences, Universit\'{e} Catholique de Louvain, 1348 Louvain-la-Neuve, Belgium}
\author{G. Hautier}
\email{geoffroy.hautier@uclouvain.be}
\affiliation{Nanoscopic Physics, Institute of Condensed Matter and Nanosciences, Universit\'{e} Catholique de Louvain, 1348 Louvain-la-Neuve, Belgium}

\date{\today}

\begin{abstract}
%One or two sentences providing a basic introduction to the field, comprehensible to a scientist in any discipline. 
The search for new materials, based on computational screening, relies on methods that accurately predict, in an automatic manner, total energy, atomic-scale geometries, and other fundamental characteristics of materials. Many technologically important material properties directly stem from the electronic structure of a material, but the usual workhorse for total energies, namely density-functional theory, is plagued by fundamental shortcomings and errors from approximate exchange-correlation functionals in its prediction of the electronic structure.
%Two to three sentences of more detailed background, comprehensible to scientists in related disciplines.
At variance, the $GW$ method is currently the state-of-the-art {\em ab initio} approach for accurate electronic structure. It is mostly used to perturbatively correct density-functional theory results, but is however computationally demanding and also requires expert knowledge to give accurate results. 
Accordingly, it is not presently used in high-throughput screening: fully automatized algorithms for setting up the calculations and determining convergence are lacking.
%One sentence clearly stating the general problem being addressed by this particular study.
In this work we develop such a method and, as a first application, use it to validate the accuracy of  $G_0W_0$ using the PBE starting point, and the Godby-Needs plasmon pole model ($G_0W_0^\textrm{GN}$@PBE), on a set of about 80 solids.
%One sentence summarizing the main result (with the words “here we show” or their equivalent).
The results of the automatic convergence study utilized provides valuable insights. 
Indeed, we find  correlations between computational parameters that can be used to further improve the automatization of $GW$ calculations. Moreover, we find that  $G_0W_0^\textrm{GN}$@PBE shows a correlation between the PBE and the $G_0W_0^\textrm{GN}$@PBE gaps that is much stronger than that between $GW$ and experimental gaps. However, the $G_0W_0^\textrm{GN}$@PBE gaps still describe the experimental gaps more accurately than a linear model based on the PBE gaps.
%Two or three sentences explaining what the main result reveals in direct comparison to what was thought to be the case previously, or how the main result adds to previous knowledge.
With this work we hence show that $GW$ can be made automatic and is more accurate than using an empirical correction of the PBE gap, but that for accurate predictive results for a broad class of materials an improved starting point or some type of self-consistency is necessary. 
%One or two sentences to put the results into a more general context. 
%Two or three sentences to provide a broader perspective, readily comprehensible to a scientist in any discipline, may be included in the first paragraph if the editor considers that the accessibility of the paper is significantly enhanced by their inclusion. Under these circumstances, the length of the paragraph can be up to 300 words. (The above example is 190 words without the final section, and 250 words with it).
\end{abstract}

\maketitle

%\end{frontmatter}

%\bibliographystyle{elsarticle-num}

\section{Introduction}
Fifty years after its formal introduction by Hedin,\cite{hedin65} and thirty years after the first applications to 'real' solids,\cite{louie85prl, Hybertsen:prb86, godby86prl} the $GW$ method has become the standard approach for studying the electronic structure of solids. In various flavors, ranging from single shot $G_0W_0$ to fully self-consistent, it is included in many multi-purpose first-principles codes.\cite{vaspgw1, yambo, berkeleygw12, aimsgw,
abinit, abinit16, exciting, schindlmayer10, vanschilfgaarde04prl, vansetten13, gwspacetimegodby, west, sax, Kutepov/etal:2012, Jiang/etal:2013} 

Despite this long maturation period, performing even the simplest $G_0W_0$ calculation, not even considering the choice of starting point,\cite{Holm/vonBarth:2004,rinke05njp,Fuchs/etal:2007,Koerzdoerfer/Marom:2012,caruso12prb,Marom/etal:2012,bruneval13jctc,Atalla/etal:2013,dauth/2015,kaplan16} on a single solid is not a trivial exercise in any of the above mentioned implementations. The complications appear at different levels. 
\begin{itemize}
\item $GW$ has a scaling of the computational complexity with system size (as represented by its number of electrons $N$) that is in common implementations at best $\mathcal{O}(N^4)$. Setting the computational parameters concerning memory and number of CPUs is hence much more involved than in typical Density Functional Theory (DFT) calculations (with a scaling on $\mathcal{O}(N^3)$ or $\mathcal{O}(N^2 \ln N)$).
\item $GW$ shows a slower convergence with respect to the basis-set size as compared to DFT or Hartree Fock (HF). This worse convergence is linked to the need to accurately describe the cusp in the electron-electron pair correlation function, similar to the convergence of the Random Phase Approximation (RPA) total energy methods.\cite{RPAreview, gulans14} Consequently, the results are also more sensitive to the quality of pseudo-potentials (PP),\cite{dixit:2010} projector augmented wave (PAW) data sets,\cite{klimes14} and resolution of the identity (RI) auxiliary basis sets,\cite{vansetten15gw100} and the choice of local orbitals in full-potential linearized augmented-plane-wave (FLAPW) approaches.\cite{Friedrich06}
\item $GW$ calculations introduce additional, in some cases interlinked, computational parameters that need to be tested for convergence.\cite{stankovski:2011,gao16}
\item In most implementations a $GW$ calculation is a 3-4 step process where the various input and output files need to be linked. 
\item Systematic validation studies using multiple codes and a wide variety of systems have recently been performed for solids at the DFT level,\cite{Lejaeghereaad3000} and for $GW$ for molecules.\cite{vansetten15gw100, scGW100, GW100vasp} For $GW$ calculations on solids validation studies usually however restrict to one code and only a limited amount of systems.\cite{vanschilfgaarde06, xin12, klimes14, nabok16} 
A systematic evaluation of the accuracy of the method is hence tedious at best.
\end{itemize}

Considering these difficulties and the lower general familiarity with the method, especially in the context of applications in the increasingly popular field of computational materials design and for high-throughput screening calculations, there is clearly a need for new automated approaches to perform $GW$ calculations. Typical automatization schemes for DFT usually rely on over-converged safe computational settings or heuristic approaches.\cite{Jain2013,oqmd,Curtarolo2012} For $GW$ calculations especially the worse scaling of the computational cost and memory do not allow for such an approach for all parameters. Hence, an automatic scheme is required to perform individual convergence studies for each compound.  

In this paper we introduce such a framework for automatic $GW$ calculations and apply it as a first example to about 80 solids establishing the accuracy and convergence properties of $G_0W_0^\mathrm{GN}$@PBE: single shot perturbative $G_0W_0$ starting from DFT results using the Perdew-Burke-Ernzerhof (PBE)\cite{tmf11} exchange correlations functional and using the Godby-Needs Plasmon-Pole model\cite{godby:prb88} (GNPPM) for the response function. It should be clear that choosing single shot $G_0W_0$, i.e. neglecting any form of self-consistency, PBE as a staring point and GNPPM as an approximation to the full RPA response function poses a limitation on the conclusions that can be drawn from the benchmark part of this manuscript. On top of that, the usage of the norm-conserving pseudopotentials, in contrast to full potential, all electron calculations introduces an approximation whose effects are currently still under debate. Removing these four uncertainties, however, is beyond the goal of the current paper. Here we hence establish the methodology needed for such future works and make a first step for one of the simplest and most controllable flavors of $GW$.

We show that in our $G_0W_0^\mathrm{GN}$@PBE results there are correlations between the values certain computational parameters need to take to reach convergence. This information can be used to further improve the automatization. From the validation study we learn that there is a strong correlation between the error of the $G_0W^\mathrm{GN}_0$@PBE gap (with respect to the experimental gap) and the experimental values. We also observe that this correlation is different for materials with or without transition metals. Moreover, we find that the correlation between the PBE gap and the $G_0W^\mathrm{GN}_0$@PBE gap is stronger than the correlation between the $G_0W^\mathrm{GN}_0$@PBE and the experimental gaps. In other words, the average error made by approximating the $G_0W^\mathrm{GN}_0$@PBE gap from the PBE gap using the linear relation between them is smaller than the average error of $G_0W^\mathrm{GN}_0$@PBE in reproducing the experimental gap. Finally, and most importantly, we show that $G_0W^\mathrm{GN}_0$@PBE still outperforms a linear model trained on the experimental gaps, i.e., even in its simplest flavor, $GW$ predicts the experimental gaps more accurately than an empirical correction based on the PBE gaps.

\section{Methodology}

In its most common approximation, which we adopt in this work, the $GW$-method provides corrections to the Kohn Sham (KS) band structure via the linearized quasi particle equation
\begin{equation}
\epsilon^{\mathrm{QP}}_{n\kk} = \epsilon^{\mathrm{KS}}_{n\kk} + \mathrm{Z}_{n\kk} \left< {n\kk}\left|  \Sigma\left(\epsilon^{\mathrm{KS}}_{n\kk}\right) - \mathrm{V}_\mathrm{xc}\right| {n\kk}\right>,
\end{equation}
where $\epsilon^{\mathrm{QP}}$ and  $\epsilon^{\mathrm{KS}}$ are the quasiparticle  (QP) and KS energies respectively, $\mathrm{V}_\mathrm{xc}$ is the exchange correlation potential, $\Sigma\left(\epsilon\right)$ the $GW$ self energy and 
\begin{equation}
\mathrm{Z}_{n\kk} =  \left(1-\left< {n\kk}\left|  \left.\frac{\partial\Sigma\left(\epsilon\right)}{\partial\epsilon} \right|_{\epsilon^{\mathrm{KS}}_{n\kk}} \right| {n\kk}\right> \right)^{-1},
\end{equation}
a renormalization factor originating from the linearization procedure. Adopting the usual decomposition of $\Sigma$ in the energy independent exchange and energy dependent correlation part,
\begin{equation}
\Sigma\left(\epsilon\right)=\Sigma_\mathrm{x} + \Sigma_\mathrm{c}\left(\epsilon\right),
\end{equation}
we have the following expressions for the diagonal matrix elements
\begin{equation}
\label{eq:diag_mat_el_Sigma_x}
 \left<n\kk\left|\Sigma_\mathrm{x}\right|n\kk\right> =
 -\frac{4\pi}{V} \sum_{\nu}^\occ \sum_\qq^\BZ \sum_{\GG}^{\Omega_\mathrm{x}}
  \frac{\abs{M_\GG^{n\nu}(\kk,\qq)}^2}{\abs{\qG}^2}
\end{equation}
\begin{equation}
\begin{split}
\label{eq:mat_el_Sigma_c}
\left<n\kk\left|\Sigma_\mathrm{c}(\omega)\right|n\kk\right> =
\frac{i}{2\pi V}
\sum_\qq^\BZ 
\sum_{\GG_1\GG_2}^{\Omega_\mathrm{c}}
\sum_{m=1}^{N_\mathrm{b}}
\Bigl[M_{\GG_1}^{mn}(\kk,\qq)\Bigr]^\dag \\ M_{\GG_2}^{mn}(\kk,\qq)\,v_{\GG_1 \GG_2}(\qq)\,
J_{\GG_1\GG_2}^{\,m \kk-\qq}(\qq,\omega),
\end{split}
\end{equation}
where $v$ is the Coulomb potential in Fourier-space and  $\Omega_\mathrm{x/c}$ defines the sphere in G space for the exchange and correlation parts of $\Sigma$ respectively: $\GG\in\Omega_\mathrm{x/c}$ if $\frac{\abs{\qG}^2}{2} \le E_\mathrm{c/x}$. In the remainder of this paper these two cutoffs will be referred to as $E_\mathrm{c}$ and $E_\mathrm{x}$ for the correlation and exchange part cutoffs respectively.\footnote{In abinit these cutoffs are controled via the input variables \texttt{ecutsigx} and \texttt{ecuteps} respectively.} $N_\mathrm{b}$ denotes the number of KS-orbitals used in constructing $G_0$ and $W_0$, the same value is also used in constructing $J$. In the exact formulation, all occupied and infinitely many unoccupied single particle states would need to be included. In practice, the total number of available states is truncated by the finite number of basis functions for the single particle states. In constructing $\Sigma$, it is further constrained by $N_\mathrm{b}$. Determining the values for $E_\mathrm{c}$ and $N_\mathrm{b}$ that lead to converged results is one of the most important steps in a $GW$ convergence study. The matrix elements $M$ are given by
\begin{eqnarray}
\label{eq:def_oscillator}
M^{b_1b_2}_\GG (\kk,\qq) & \equiv& \left<\kk-\qq,b_1|e^{-i(\qq+\GG)\cdot\rr}|\kk,b_2\right> \\
& = & \sum_{\GG'} u_{\kk-\qq b_1}^\dagger(\GG')u_{\kk b_2}(\GG+\GG'). \nonumber
\end{eqnarray}

The matrix elements $J$  in Eq.~\ref{eq:mat_el_Sigma_c} originate from the screening and take a particular form depending on the approach used to describe the frequency dependency of $W$. In this work, all calculations are performed using the Godby-Needs plasmon pole approximation\begin{equation}
\label{eq:ppmodel_real}
\Re\,\epsilon^{-1}_{\GG_1 \GG_2} (\qq,\omega) =
\delta_{\GG_1 \GG_2} + \dfrac{\Omega_{\GG_1\GG_2}^2(\qq)}{\omega^2-\tilde\omega^2_{\GG_1\GG_2}(\qq)}
\end{equation}
in which the parameters $\tilde\omega$ and $\Omega$ are derived to reproduce the ab initio inverse dielectric matrix $\epsilon^{-1}$ computed at the static limit and at an additional imaginary frequency point.\cite{abinit16,abinit,godby:prb88,godby86prl}
In terms of $\tilde\omega$ and $\Omega$, the matrix elements of $J$ are given by:\cite{giantomassi11,giantomassi-thesis}
\begin{equation}
\begin{split}
J^{m \kk-\qq}_{\GG_1\GG_2}(\qq,\omega) = \Omega_{\GG_1\GG_2}^2(\qq) \\
 \int \dfrac{e^{i\omega'\delta} \,\mathrm{d} \omega'}{\Bigl(\omega+\omega'-\epsilon_s +i\eta\,\mathrm{sign}(\epsilon_s-\mu)\Bigr)\Bigl(\omega'^2-(\tilde\omega_{\GG_1\GG_2}(\qq)-i\eta)^2\Bigr)}
\end{split} 
\end{equation}

All calculations presented in this work are performed using the ABINIT software package,\cite{abinit} employing newly developed Optimized norm-conserving Vanderbilt
pseudopotentials (ONCVPSP).\cite{oncvpsp} Two projectors per angular momentum channel are used, in order to include semi core states while at the same time keeping a good description of the continuum states in the empty region, both of which are crucial for accurate $GW$ calculations. In general we observe a close agreement of the logarithmic derivatives up to energies of 300~eV. Deviations of the logarithmic derivative of the pseudized wavefunctions from that of the reference atomic all-electron calculation are an indication of possible ghost states. Manual inspection of the band structures of elemental solids up to 100 eV above the Fermi level did not reveal any ghost state in the ONCVPSP pseudo-potentials used in this work.\cite{oncvpsp} 

Our automatic $GW$ workflow consists of the following steps:  

\begin{enumerate}
\item Convergence testing of the KS energies with respect to the basis set, i.e., the energy cutoff of the plane wave basis; 
\item Convergence testing of the QP energies with respect to the number of empty states, $N_\mathrm{b}$, and cutoff in the correlation part of $\Sigma$,  $E_\mathrm{c}$, at a low density k-mesh;
\item Testing of the convergence behavior of the QP energies at a high density k-mesh;
\item Post processing of the results calculated at the high density k-mesh: constructing a scissor operator and band structure and storing it in a database.
\end{enumerate}

The flow itself, its generation, execution, and post processing are programmed within the AbiPy python framework.\cite{abipy, abinit16} AbiPy is an open-source library for the analysis of the results produced by ABINIT based on the Python ecosystem powering the Materials Project, Pymatgen.\cite{pymatgen} The following sections describe and validate these steps in detail.
 
\subsection{Convergence at the KS level}

In the automatic flow the energy cutoff for the wavefunctions is fixed at 44~Ha (1197~eV) and an automatic test up to 52~Ha (1415~eV) is conducted to ensure convergence. For none of the compounds the convergence of the total KS eigenvalue energy range\footnote{The difference between the highest (unoccupied) and lowest (occupied) KS eigenvalue included in the calculations.} indicated that a larger value was needed. The other, ground-state related, computational parameters are fixed in the flow to ensure convergence is reached within 0.05 eV on the full KS band width.

\subsection{Automatic convergence testing at the QP level}

To optimize the performance of the entire flow, it is essential to decouple those parameters for which the convergence can be studied independently. In general the energy cutoff of the correlation part of the self-energy $\Sigma$  ($E_\mathrm{c}$) and the number of empty states ($N_\mathrm{b}$) are coupled and need to be considered simultaneously. The convergence properties of this pair and that of the k-mesh density are, however, decoupled. The value of $E_\mathrm{x}$ is fixed to the same 44~Ha as used for the KS calculation. 

The convergence with respect to $E_\mathrm{c}$ and $N_\mathrm{b}$ is studied as a 2D problem at a low density ($\Gamma$ centered $2\times2\times2$) k-point mesh. On a $4\times4$ ($E_\mathrm{c}$ $\times$ $N_\mathrm{b}$) grid of parameters, a full $GW$ calculation is performed. First, for fixed $E_\mathrm{c}$, the converged $N_\mathrm{b}$ values are determined by fitting an asymptotic function. To improve the stability of the fitting procedure, the algorithm fits multiple functions with only two parameters and selects the best fit.\footnote{Reciprocal functions with powers in the range of 0.5 to 6 are fitted to the data. The best fit for each power is compared, the one with the best overall agreement is used to approximate the asymptotic value.} Second, the final converged result is obtained by fitting the converged results at fixed $E_\mathrm{c}$ obtained in the previous step. This final value is used to determine for which pair ($E_\mathrm{c}$, $N_\mathrm{b}$) the result is within a predefined distance from the converged result. The pair ($E_\mathrm{c}$, $N_\mathrm{b}$) is chosen to minimize $E_\mathrm{c}$ in order to reduce the computational cost. If no ($E_\mathrm{c}$, $N_\mathrm{b}$) pair can be determined on the current grid, the grid is automatically extended to add more data-points until the converged pair is found.

\begin{figure}[!ht] 
\centering
\includegraphics[width=1.0\columnwidth]{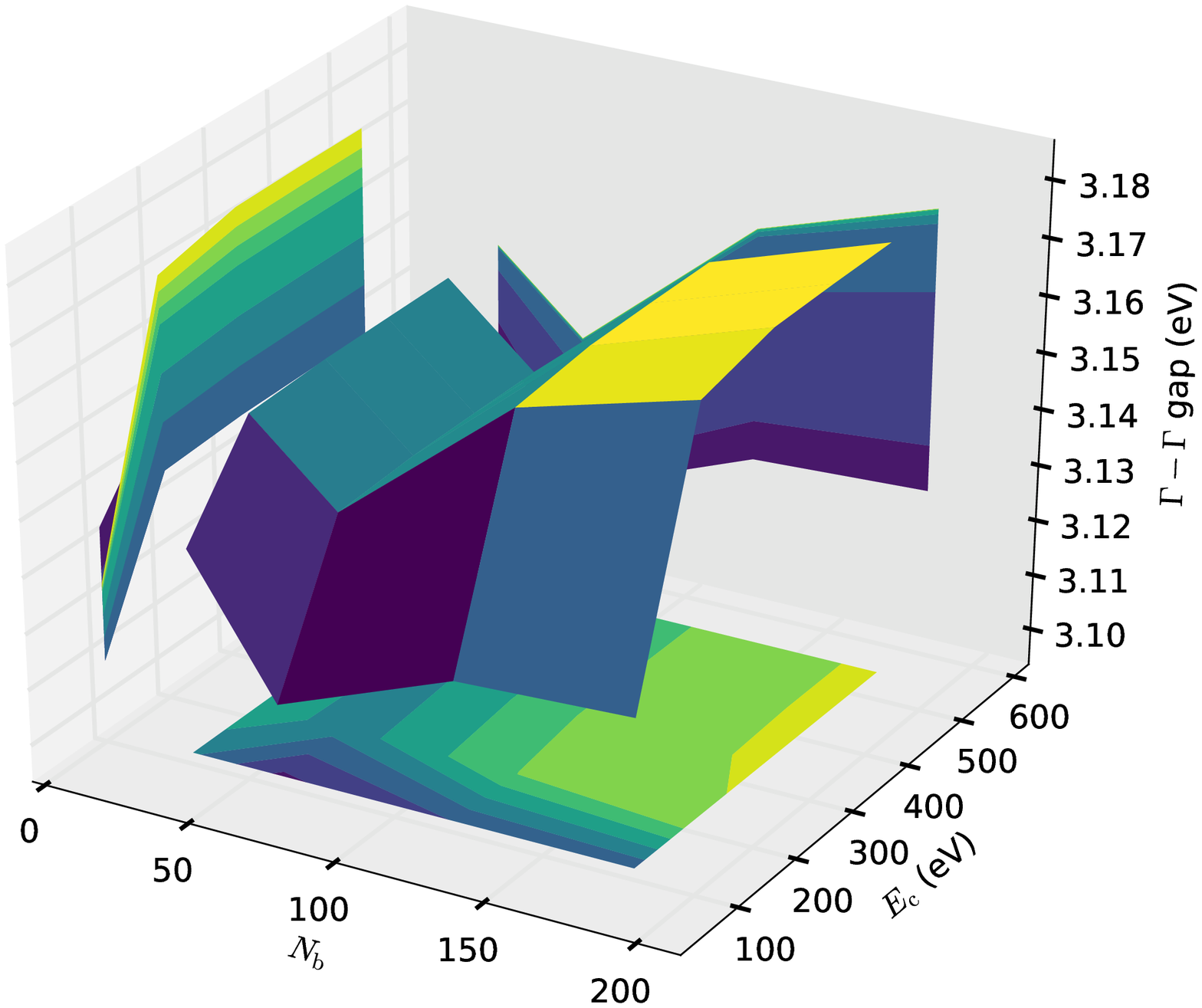} 
\includegraphics[width=1.0\columnwidth]{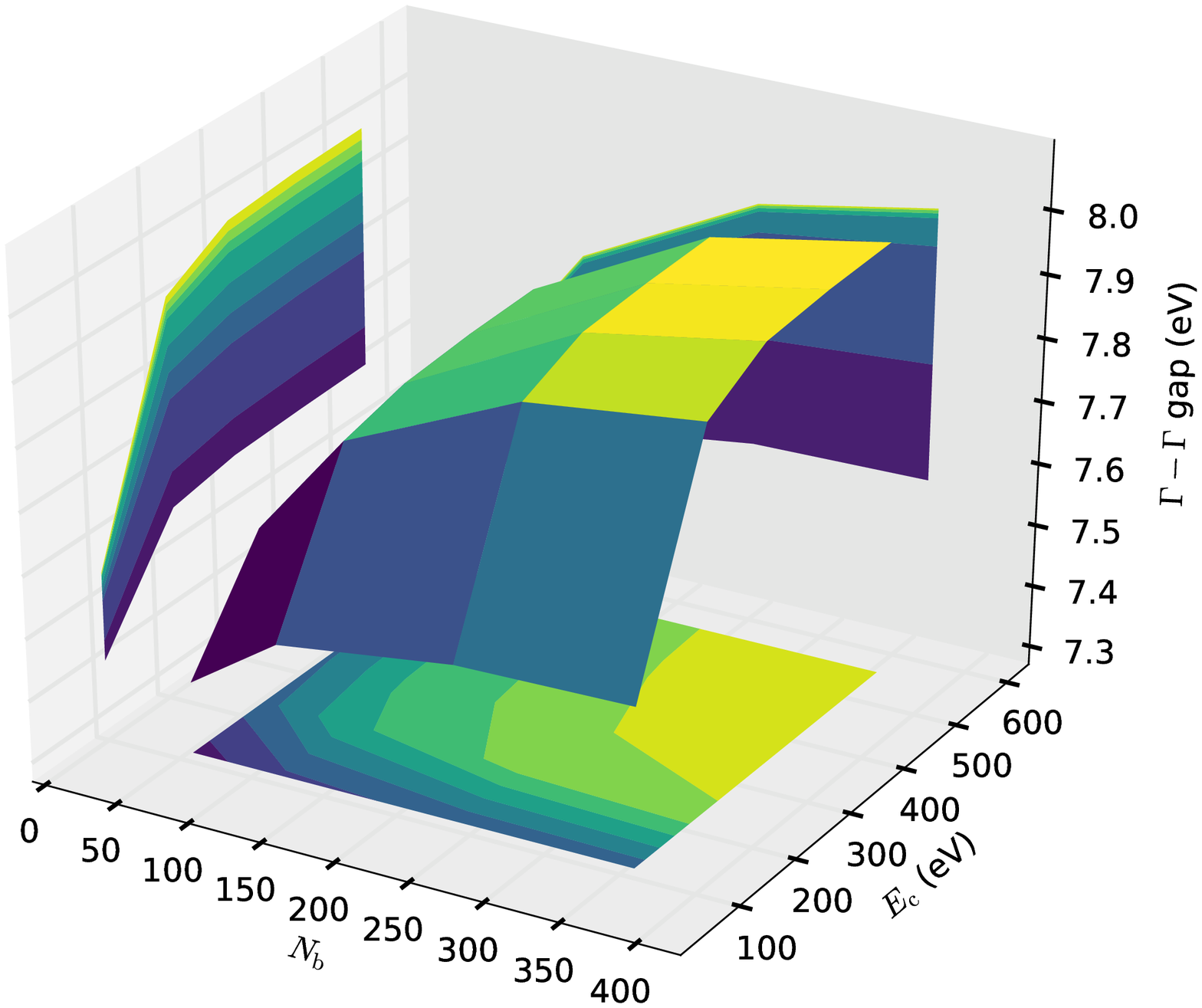} 
\caption{The convergence of the gap at $\Gamma$ as a function of the number of bands ($N_\mathrm{b}$) and the cutoff used for the screening and $\Sigma$ ($E_\mathrm{c}$) for Si (upper) and BN (lower).}\label{plot-2dconv}
\end{figure}

Figure~\ref{plot-2dconv} shows two examples of the convergence of the fundamental QP-gap at $\Gamma$ with respect to $E_\mathrm{c}$ and $N_\mathrm{b}$ for silicon and boron nitride. Both Si and BN show the typical coupled convergence behavior but the convergence rate is an order of magnitude larger for BN. This difference already shows that employing heuristic rules to perform automatic GW calculations in several systems would be error-prone since the convergence rate is strongly system-dependent.  Rules of thumb extracted from a restricted dataset hence cannot be extrapolated to a large class of systems. The coupled behavior between $E_\mathrm{c}$ and $N_\mathrm{b}$ will be discussed in detail in sect.~\ref{sect:convparcor}.
 
\begin{figure}[!ht]
\centering
\includegraphics[width=0.9\columnwidth]{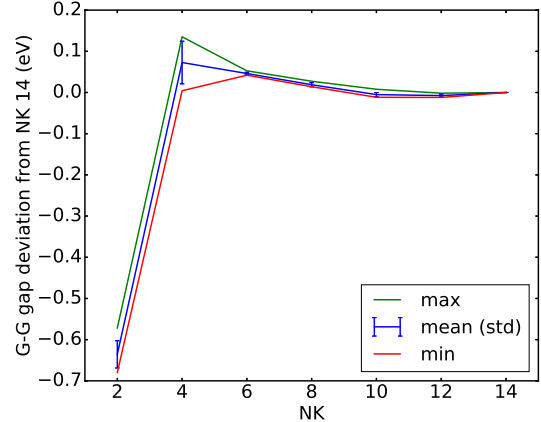}
\caption{The ($E_\mathrm{c}$, $N_\mathrm{b}$) convergence data summarized for 7 k-point meshes, labelled by NK, corresponding to a NK x NK x NK grids. For each k-mesh the average, maximum and minimum deviation from the data at the largest density grid ($14\times14\times14$) is plotted. The data is shown for gold.}\label{plot-ecuteps-kpoint-decoupling}
\end{figure}

The decoupling of the convergence with $E_\mathrm{c}$ and $N_\mathrm{b}$ and the density of the k-point mesh is shown in Fig.~\ref{plot-ecuteps-kpoint-decoupling}. Here the ($E_\mathrm{c}$, $N_\mathrm{b}$) convergence data is summarized for 7 k-point meshes for Au.\footnote{In calculating the response function for metals no drude contribution is included.} For this example a metal was chosen for which the dependence on the k-mesh is larger than for the semi conductors and insulators used in this study. Because of this no Drude contribution is included in these specific calculations; the example is only used to show that the decoupling takes place even in the case of metallic occupation. For each k-mesh we plot the average, maximum and minimum deviation from the results obtained with the highest density grid. The average shows how well the actual k-mesh gives converged results, while the difference between minimum and maximum deviation measures how well the shape of the convergence surface is described by a specific k-mesh. The small difference between the minimum and maximum already with the $2\times2\times2$ mesh shows that performing the convergence study using this coarse grid is valid. 

\begin{figure}[!ht]
\centering
\includegraphics[width=0.9\columnwidth]{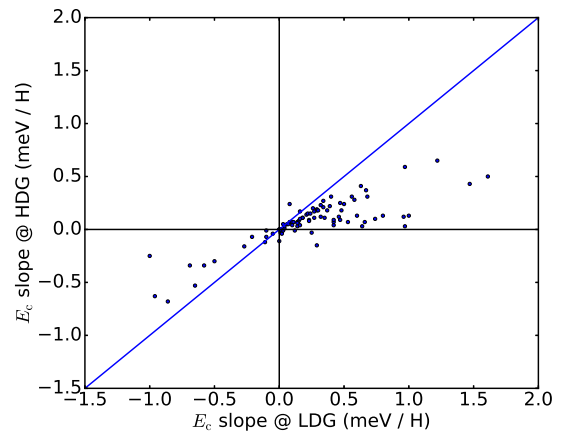}
\caption{Relation between the $E_\mathrm{c}$ slopes using the Low Density Grid (LDG) and High Density Grid (HDG)\label{fig-es}}
\end{figure}

\begin{figure}[!ht]
\centering
\includegraphics[width=0.9\columnwidth]{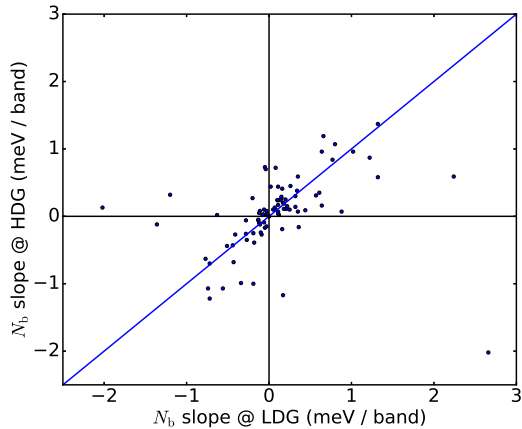}
\caption{Relation between the $N_\mathrm{b}$ slopes using the Low Density Grid (LDG) and High Density Grid (HDG)\label{fig-bs}}
\end{figure}

The final computational step consists of a test to ensure that the convergence rate at the high-density k-point grid (HDG) is indeed similar to the one at the low-density grid (LDG). To this end, four full $GW$ calculations are performed at a predefined HDG. From these four data sets, the derivative of the gap with respect to $E_\mathrm{c}$ and $N_\mathrm{b}$ is calculated by means of finite differences and compared to the corresponding values calculated at the LDG. In about 90\% of the systems studied here, the $E_\mathrm{c}$ slopes at the HDG are actually lower than on the LDG. For the remaining systems, the $E_\mathrm{c}$ slopes are only marginally larger at the HDG, see Fig.~\ref{fig-es}. For $N_\mathrm{b}$ the situation is less clear. However, the cases where the slopes are larger than 1 meV at the LDG do decrease in the HDG. For the smaller slopes we observe a more scattered behavior, see Fig.~\ref{fig-bs}. Hence for all systems in this set, the converged parameters found on the LDG can be safely used on the HDG to obtain converged results. A full list of the numerical data is available in the supplemental material.

\subsection{Postprocessing}
\label{postp}
\begin{figure}[!ht]
\centering
\includegraphics[width=0.9\columnwidth]{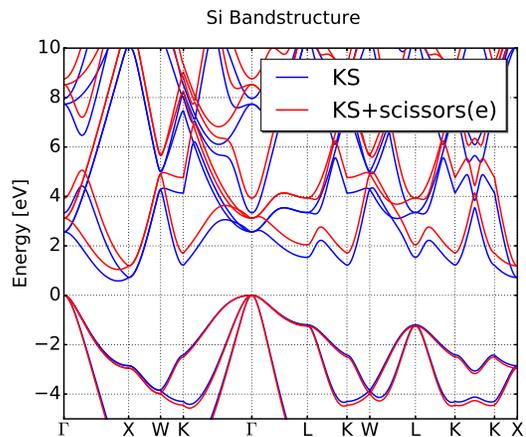}
\caption{The band structure of silicon computed at the KS level and with the energy dependent extrapolation function described in the text.}\label{plot-sibands}
\end{figure}

After the final converged calculation is performed the flow performs various post processing steps. The output of the final $GW$ calculation is stored in a database, as a queryable entry containing a link to a simultaneously stored NetCDF output file.\cite{abinit16} For the full QP-spectrum a bilinear energy dependent extrapolation function is constructed. For this an improved version of the polyfit method\cite{waroquiers13} is used. The first linear section passes through the valence band maximum (VBM) correction and optimally fits the corrections to the other occupied QP states. The second section passes through the conduction band minimum (CBM) correction and optimally fits the corrections to the other unoccupied states.\footnote{This method of extrapolation has the advantage of being numerically robust; For a given set of corrections a unique extrapolation function is always found. It has, however, the disadvantage that it breaks down in the cases in which DFT incorrectly predicts a metal, e.g. CdO, GeS, Mg$_2$Sn. The development of an approach that would not have this problem and is robust enough to be used in a fully automatic fashion is still in progress.} The extrapolation function is then used to perform corrections to a separately calculated full band structure. Figure~\ref{plot-sibands} shows the band structure of silicon obtained using this procedure, with the bands aligned at the valence band maximum. 

In the analysis of the complete data set we look for relations between gaps at different levels of theory, converged input parameters, and other a priori known quantities. T-tests are used to establish whether the obtained results are statistically relevant. Regression obtained using Ordinary Least Square (OLS) minimization and using Robust Linear Models (RLM) using the Huber T norm,\cite{hubertnorm}  are both used to obtain the sought after relations. In the latter, the deviations $d$ weighted according to:
\begin{eqnarray}
w(d)  & = & 1           \; ; \;\;  d \le t \nonumber\\
w(d)  & = & \frac{t}{d};   \;\; d > t
\end{eqnarray}
using the default value of $t=1.345$, are minimized instead of the bare deviations in the OLS. It is therefore less sensitive to outliers than OLS and comparing the two approaches gives an indication of the presence of real outliers.

\section{Results for $G_0W^\mathrm{GN}_0$@PBE.}

As a first application, we use the above-described methodology on a set of solids used previously for evaluating various DFT functionals and the gap-prediction method $\Delta-sol$,\cite{chan10} using the Godby-Needs plasmon-pole model as implemented in \textsc{abinit}.\cite{abinit16,abinit,godby:prb88} Of this original set those systems are included that are not metallic at the KS level and have a size that allowed a converged $G_0W^\mathrm{GN}_0$@PBE given our current computational resources. In the final set, there are three elemental compounds and seventy-five binaries with stoechiometry occurences fifty-one XY, thirteen XY$_2$, six XY$_3$, one XY$_4$, three X$_2$Y$_3$ and one X$_2$Y$_5$. Among the binaries there are twelve III-V , twenve II-VI, eleven I-VII, and eight IV-VI compounds. Anions may also serve to characterise our set. There are twelve halides, thirty-seven chalchogenides (including twelve oxydes), and twenty pnictides (including four nitrides). The full results for all systems included in this study\footnote{\systems} are given in the Supplemental Material. A comparison will be made in this section to experimental gaps collected in Ref~\citenum{chan10}. To make the comparison in a proper way there are however three different effects that should be taken into account.
 
First, the experimental values have all been measured at room temperature. As such they have a tendency to underestimate the purely electronic gap, at zero Kelvin without zero-point renormalization, by 0.1 to 0.4~eV.\cite{chang86,manjon04,cardona05,giustino10,antonius15,ponce15} For 31 of the systems presented in this section, we could obtain a 0K extrapolated band gap using literature values.\cite{gappaper} For these we find an average underestimation of 0.12~eV with a median of 0.09~eV. Second, the calculated results only include relativistic effects at the scalar relativistic level. For systems containing heavy elements, the inclusion of spin orbit corrections can lead to reductions of the band gap by up to 0.5 eV.\footnote{A spinor version of $GW$ is currently under development in ABINIT.} \cite{scherpelz16} Finally, in the convergence studies the convergence criterion was set to 0.1 eV. Since in almost all cases the asymptotic value band gap is approached from below in general we underestimate the fully converged result by up to 0.1~eV. Taking the cumulative effect of these three effects into account leads to a window of 0--0.9 eV. This means that even the (scalar relativistically) exact electronic structure theory would overestimate the room temperature experimental gap by 0--0.9 eV. In comparisons made below, this interval is indicated by two blue lines, i.e., `perfect agreement with experiment' would lead to a datapoint between the two blue lines. 
 
Besides these three sources of errors originating from approximations in the computational setup, the experimental results carry an uncertainty. For about half of the systems presented in this section we have found additional experimental room temperature (290 - 300 K) gaps in the literature.\cite{gappaper} Within these collected results we find an average standard deviation of 0.16 eV, with a median of 0.08 eV. 

The full data, details about all systems considered in this set, all numerical values, and full detail on the statistical analysis are available in the supplemental material. 

\subsection{Evaluation of the accuary of $G_0W_0^\textrm{GN}$@PBE}

\begin{figure}[!ht]
\centering
\includegraphics[width=0.90\columnwidth]{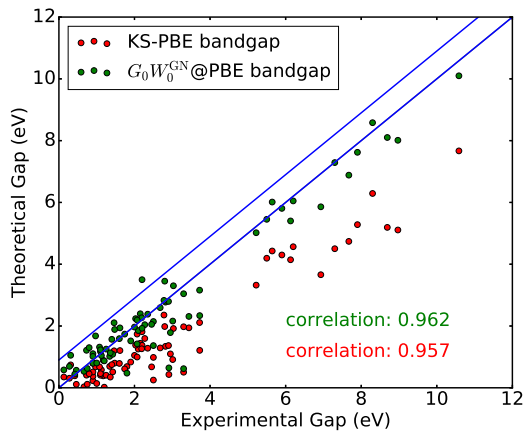}
\caption{Comparison of the KS and QP fundamental gaps evaluated from a band structure computed on a high-symmetry path through the Brillouin zone. The QP band structures are obtained from applying the extrapolation function to the KS band structure. The blue lines indicate the estimated interval in which exact agreement would be expected taking spin-orbit coupling, finite temperature, and the level of convergence into account, see text.}\label{plot-bgaps}
\end{figure}

%The  $G_0W_0^\textrm{GN}$@PBE band gaps calculated on the regular k-point grid are compared to the experimental values and their KS counterparts in Fig.~\ref{plot-fgaps}.  
In Fig.~\ref{plot-bgaps} the calculated gaps are obtained from a calculation on high symmetry lines through the Brillouin zone. The  $G_0W_0^\textrm{GN}$@PBE results are obtained from correcting the KS band structure using an energy-dependent extrapolation function, as described in section~\ref{postp}. 

%\begin{figure}[!ht]
%\centering
%\includegraphics[width=0.9\columnwidth]{plot_fgaps.eps}
%\caption{Comparison of the KS-PBE and $G_0W_0^\textrm{GN}$@PBE fundamental gaps on the regular mesh. The correlation between the $GW$ and experimental gaps is \gwexcor. The blue lines indicate the estimated interval in which exact agreement would be expected taking relativistic effects, finite temperature, and the level of convergence into account, see text.
%}\label{plot-fgaps}
%\end{figure}
 
In general, we observe the typical opening of the gap by $G_0W_0^\textrm{GN}$@PBE as observed in many previous studies for various flavors of $GW$.\cite{hedin65, Hybertsen:prb86, aryasetiawan, vanschilfgaarde06, klimes14,larson2013} However, we do observe a much larger spread of the $GW$ error for this set of systems. 

\begin{figure}[!ht]
\centering
\includegraphics[width=0.90\columnwidth]{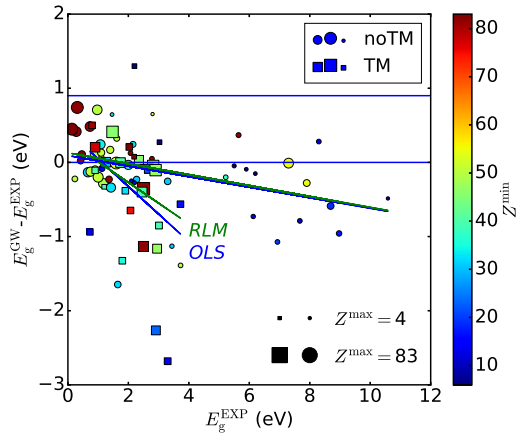}
\caption{The error of the GW gap with respect to the experimental value compared to the experimental value. The squares indicate compounds containing transition metals, the circles compound without transition metals. The size of the symbols reflects the smallest atomic number occurring in the compound, the color the largest atomic number. The horizontal blue lines indicate the estimated interval in which exact agreement would be expected taking spin-orbit coupling, finite temperature, and the level of convergence into account.}\label{plot-diff}
\end{figure}

The comparison made in Fig.~\ref{plot-bgaps} is reevaluated in Fig.~\ref{plot-diff} subtracting the experimental values and adding various identifiers, identifying which compounds contain transition metals and the mass of the lightest and heaviest element present.  Zero point renormalization and relativistic effects (not take into account here) might be the origin of larger errors in systems with light and heavy elements, respectively. However, the identification of light and heavy element systems in Fig.~\ref{plot-diff} shows that this is not the dominating effect.

Making the distinction on the presence of transition metal elements (elements for column 3 to 12 of the periodic table), on the other hand, reveals a clear correlation of the $G_0W_0^\textrm{GN}$@PBE error with the experimental gap. However, the relation, for an ordinary least square (OLS) and robust linear model (RLM) using the Huber T norm with median absolute deviation scaling,\cite{hubertnorm} is different between the two groups. For the transition metal compounds we have \TMfunc\; and \TMRLMfunc . For the non-transition metal compounds we have \nonTMfunc\; and \nonTMRLMfunc . The trends observed here agree well with Lany's observations on 3d-transition metal oxides.\cite{lany13} For the transition metal containing compounds we observe a mean absolute deviation of 0.64~eV from the experimental gaps. For the compounds without transition metals this reduces to 0.38~eV.
 
Table~\ref{diffstat} lists the statistical evaluation of the data presented in Fig.~\ref{plot-diff}. We observe a 0.18~eV average underestimation of the $GW$ results with respect to the room temperature experimental gaps. Adding 0.45~eV to, on average, correct for relativistic effects and finite temperature we get to a 0.63~eV underestimation.

\begin{table}
\center
\input{table_full_describe.tex}
\caption{Statistics on the deviation of the QP band-gap from the experimental gap. Systems with a zero band gap in KS-PBE have been omitted. All gaps are in eV. $Z^\mathrm{max}$ and $Z^\mathrm{min}$ report the maximal and minimal atomic number present in the compound.}\label{diffstat}
\end{table}

The compounds with a $G_0W_0^\textrm{GN}$@PBE error larger than one standard deviation (\std~eV) from the mean (\mean~eV) are listed in Table~\ref{outl}. At the top of the table, we find the compounds where $GW$ is underestimating the most. We note that a significant fraction of these compounds contains copper. Besides these we find GeS, SnO$_2$ and GaN strongly underestimating. Ge, Sn, and Ga occur in various other compounds in our set that do not stand out particularly. It could however be that in these particular cases the low lying d-levels are problematic for $GW$. In CaO and NaCl the absolute value of the error is about 1, however since the actual gaps are rather large the relative errors are much smaller than in most compounds. In the lower part of the table, we found compounds containing Te and Sn in which spin-orbit effects are expected to be strong. V$_2$O$_5$ will be discussed below.

\begin{table} 
\caption{Compounds for which the $G_0W_0^\textrm{GN}$@PBE error with respect to the room temperature experimental gap is more than one standard deviation (\std eV) from the mean deviation {(\mean eV)}. Type indicates the classification as transition metal containing (2) or not (1). All gaps are in eV. $Z^\mathrm{max}$ and $Z^\mathrm{min}$ report the maximal and minimal atomic number present in the compound.}\label{outl}
\center
\input{table_outliers.tex}
\end{table} 

\subsubsection{Specific systems}

{\bf V$_2$O$_5$} - The largest overestimation in our dataset is observed in V$_2$O$_5$ in which $G_0W_0$ gives a gap of 3.50 eV that significantly exceeds the experimental value by 1.30~eV. This material has been studied by means of $GW$ calculations previously: Lany\cite{lany13}, using ev$GW_0$@GGA+U,\footnote{eigenvalue only self-consistency in $G$ with fixed $W$ starting from generalized gradient approximation with Hubbard-U correction} found a gap of 4.69~eV and Bhandari {\em et al.}\cite{bhandari15} using qs$GW$\footnote{complete quasiparticle self-consistency} found a  gap of 4.0~eV. Both (partial) self-consistent approaches are known to enlarge the gap as compared to $G_0W_0$. Both results, in this sense, agree with our 3.50~eV $G_0W_0$ gap. Bhandari {\em et al.} hypothesize the discrepancy with respect to the experimental value to be caused mainly in terms of the effects of lattice polarization.\cite{bhandari15}

{\bf ZnO} - Wurtzite ZnO is a well known problematic system for $GW$.\cite{louie10,Usuda/Schilfgaarde:2002,Gori:2010,dixit:2010,stankovski:2011,Friedrich/etal:2011, friedrich12} The physical model, i.e. the approximation used for the response function (plasmon pole or full frequency), has a large influence on the band gap \cite{stankovski:2011}, but also the convergence with respect to the number of unoccupied states is much slower in ZnO \cite{stankovski:2011,Friedrich/etal:2011} than in other solids such as silicon. Indeed our calculations confirm this behavior. Within the computational resources defined in our algorithm, the procedure does not find a set of converged parameters before the calculations become unfeasible, given the computing capabilities available to us.\footnote{The main limitation is the amount of memory available per cpu (i.e. MPI process), since, due to implementation details, in the current implementation the {\em entire } response function needs to be kept in memory on each process.} 

\subsubsection{The correlation of KS-PBE and $G_0W_0^\textrm{GN}$@PBE gaps}

\begin{figure}[!ht]
\centering
\includegraphics[width=0.90\columnwidth]{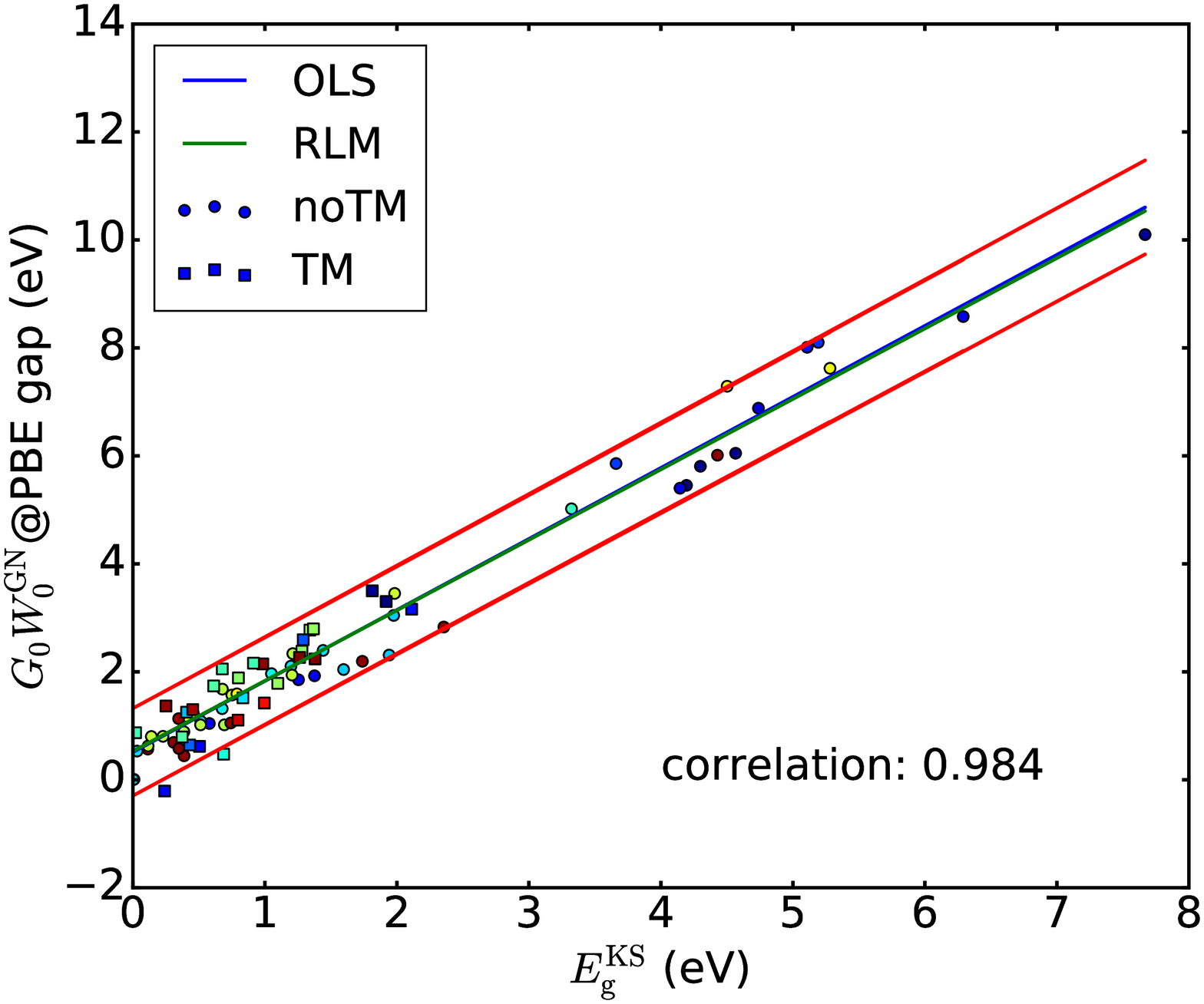}
\caption{Comparison of KS-PBE and $G_0W_0^\textrm{GN}$@PBE gaps. An Ordinary Least Squares (OLS) linear regression leads to an \qpvsksOLSfunc\; relation between the two sets with a correlation of \gwkscor. A Robust Linear Model (RLM) leads to only slightly different results: \qpvsksRLMfunc.}\label{plot-gap-corr}
\end{figure}

The relation between the KS-PBE and $G_0W_0^\textrm{GN}$@PBE fundamental gaps (on the regular k-point grid) is investigated in Fig.~\ref{plot-gap-corr}. In contrast to the comparison of the $G_0W_0^\textrm{GN}$@PBE gaps with the experimental values the data do not show a significant difference between the two groups of materials, i.e., those containing transition metals and those without transition metals. The correlation between KS-PBE and $G_0W_0^\textrm{GN}$@PBE gaps is however stronger than any other observed in the data. Moreover in performing the linear regression analysis we find remarkably little outliers; the OLS and RLM find the same parameters of the linear relation.

The relation found between the KS-PBE and $G_0W_0^\textrm{GN}$@PBE gaps can be used to estimate the QP gap directly from the KS-PBE gap. The mean absolute deviation of the estimated $G_0W_0^\textrm{GN}$@PBE gap from the actual values are 0.32, 0.37, and 0.29 eV for the whole set, transition metal containing compounds, and compounds without transition metals respectively. 

Beyond the correlations between the experimental, KS-PBE, and $G_0W_0^\textrm{GN}$@PBE gaps, the next strongest correlations are observed between the gaps and the ratio of the electro negativities of the elements present in the compounds (see section "All correlations in the dataset" in the Supplemental Material). The experimental gaps show the strongest correlation with the electronegativity ratio: 0.60. 

\subsubsection{$G_0W_0^\textrm{GN}$@PBE v.s. an empirical correction to KS@PBE }

\begin{figure}[!ht] 
\centering
\includegraphics[width=0.90\columnwidth]{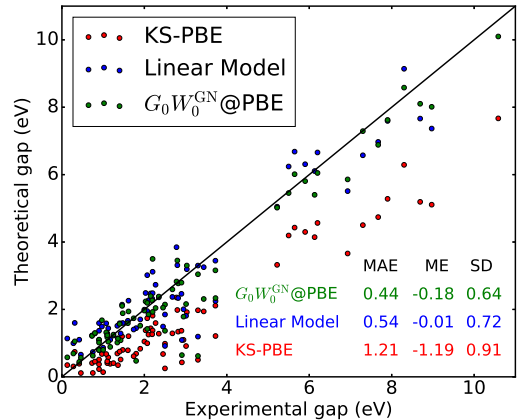}
\caption{Comparison of KS and $GW$ gaps, and gaps obtained from a five-fold crossvalidated linear model of the KS gap trained on the experimental gaps. The statistical evaluation  provides the mean absolute error (MAE), mean error (ME), and standard deviation (SD) }\label{fig-linear_model}
\end{figure}

An important question in addressing the accuracy of $G_0W_0^\textrm{GN}$@PBE is whether it  performs better or worse than would a linear model based on the KS gaps fitted to the experimental values, as proposed by Setyawan et al.\cite{Setyawan11}.  Figure~\ref{fig-linear_model} shows this comparison. The performance of KS-PBE, $G_0W_0^\textrm{GN}$@PBE and the linear model is reported in the inset. 

%\begin{table}
%\caption{Statistical evaluation (Mean Absolute Error, Mean Error, and Standard Deviation) on the comparison of the $GW$ and KS gaps and gaps obtained from a five-fold crossvalidated linear model of the KS gap trained on the experimental gaps.}\label{tab-stat-model}
%\center
%\input{ksb_lm.tex}
%\end{table}

The linear regression model is tested using five-fold cross validation. The cross validation score for the model is 0.88 (full details are provided in the supplemental material). Both the standard deviation (SD) and the mean absolute error (MAE) of the $GW$ results are significantly smaller than those of the linear model. The p-value for a paired t-test\footnote{both the LM and $GW$ errors are sufficiently well normal distributed to justify a t-test. The corresponding p-values are 0.001 and 10$^{-6}$ respectively} between the $GW$ and LM errors is 0.0006, indicating a significant difference between the two sets of results. The mean absolute error and standard deviation stay almost the same. We hence conclude that $GW$,  even the simplest form of $GW$ method (perturbative single-shot correction based on the plasmon-pole model) already significantly outperforms a single parameter empirical model fitted to the experimental results. If the linear model is trained using the $GW$ gaps the model is only better for the mean error (ME). Only when we train a model including both the type, electro negativity ratio, the nuclear mass ratio, and the KS gap we reach a model with similar accuracy as $GW$: 0.43, -0.03, and 0.59~eV for the MAE, ME, and SD respectively. 

\subsubsection{Comparison of $G_0W_0^\textrm{GN}$@PBE to $\Delta$-sol and HSE06.}

A computationally efficient method to significantly improve KS-gaps was proposed by Chan and Ceder in 2010 under the name of $\Delta$-sol.\cite{chan10} It is closely related to the $\Delta$-SCF method for finite systems where the HOMO-LUMO gap is calculated explicitly as the change in total energy upon changing the number of electrons in the system. In the $\Delta$-sol method, one electron is removed per screening volume. How many electrons effectively reside in a screening volume, $N^*$, depends on the exchange-correlation functional and on the type of system. In ref.~\onlinecite{chan10}, the values of $N^*$ are determined for systems containing $sp$ and $spd$ valence electrons for three functionals. In addition the $N^*$ are specifically geared to be predictive in a range of gaps of  between 0.5 and 4~eV. Using these parameters, mean absolute errors of 0.31~eV for a set of typical semiconductors and 0.26~eV for a larger set of transition metal containing compounds are found. The errors for gaps calculated using HSE06\cite{Heyd2003,Heyd2004,Krukau2006,Heyd2006} (0.26 and 0.41~eV) are also reported.  
The  $G_0W_0^\textrm{GN}$@PBE gaps reported in this work hence have larger mean absolute errors than the gaps obtained using $\Delta$-sol and HSE06. The 0.44~eV MAE of $G_0W_0^\textrm{GN}$@PBE, however, is achieved for systems with gaps up to 10~eV. If we only consider the 0.5~-~4.0~eV gap range for the non transition metal compounds in our set, the $G_0W_0^\textrm{GN}$@PBE MAE drops to 0.32~eV, the MAE for the compounds containing transition metals  in this range increases to 0.61~eV. Summarizing we conclude that  $G_0W_0^\textrm{GN}$@PBE has similar accuracy in predicting band-gaps as HSE06 and $\Delta$-sol for $sp$-compounds. For compounds  containing transition metals HSE06 and Delta-sol have a better accuracy.

\subsection{Convergence parameters}\label{sect:convparcor}

The significant number of $GW$ results obtained with our automatically converged studies allow us to investigate possible relationships among the values needed to reach convergence. A statistical summary of the converged parameters is given in Table~\ref{tab-convpar}. The median is for all observables clearly larger than the mean value, indicating an asymmetric distribution in all cases. 
 
\begin{table}
\center
\caption{Statistics on the convergence parameters.}
\input{table_descibe_convpar.tex}\label{tab-convpar}
\end{table}

\begin{figure}[!ht]
\centering
\includegraphics[width=0.90\columnwidth]{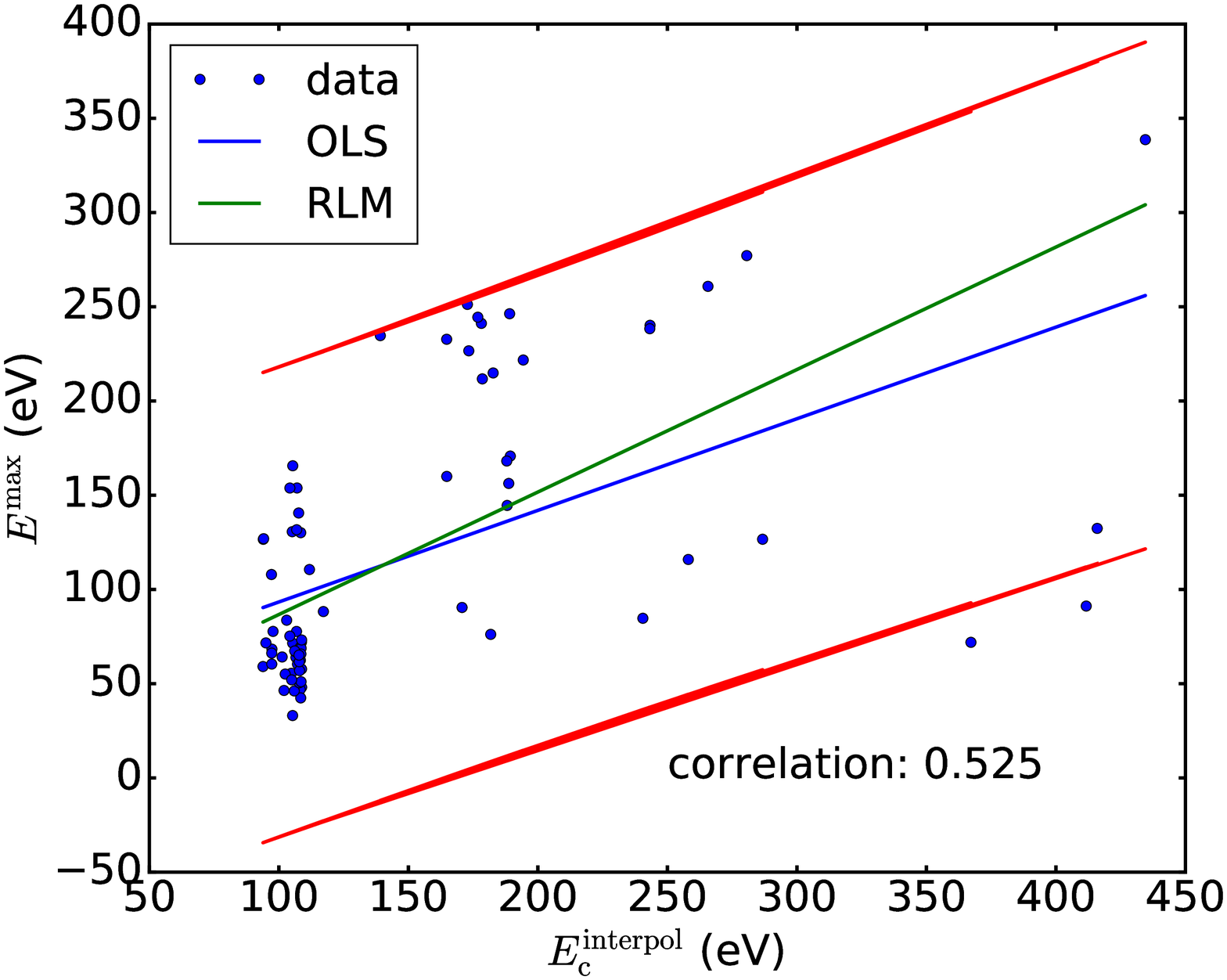}
\caption{Comparison of the energy of the highest band and the energy cutoff for $\Sigma$ needed to reach convergence.}\label{plot-mevsec} 
\end{figure}

Calculating the correlation coefficients between all quantities in our dataset reveals two significant correlations involving the convergence parameters.  The first is the correlation between the cutoff for the expansion of the screening and the self energy ($E_\mathrm{c}$) and the energy of the highest band ($E^{\mathrm{max}}$). It is investigated in Fig.\ref{plot-mevsec}. Since the grid of $E_\mathrm{c}$ values actually used is relatively coarse, the statistics is performed on an interpolated $E_\mathrm{c}$, $E_\mathrm{c}^{\mathrm{interpol}}$, which estimates the actual value for which the convergence criterium would be met. In contrast $E_\mathrm{c}$ is the first value {\em on the grid} for which it is met. The correlation between the two parameters is \ecutepsmaxencor. There is a clear difference between the OLS and the RLM fit, with the RLM looking more appropriate. Although the fit is only barely significant, the obtained results can be used to simplify convergence studies for new materials. Instead of treating the cumbersome 2D convergence problem, one could perform a convergence study on both at the same time by keeping the ratio between the two fixed.

A relation between the maximum number of bands and the cut-off energy for correlation has been observed for individual systems before,\cite{gao16,stankovski:2011} although, to our knowledge, it has not been studied in a systematic way yet. It may be rationalized as follows. The expression for the RPA dielectric matrix reveals that the $\mathbf{G}$-dependence of the screening is defined by the oscillator matrix elements (Eq.~\ref{eq:def_oscillator}) that are given by a convolution in $\mathbf{G}$-space between the periodic parts of the Bloch states associated to one occupied and one empty state. As a consequence, the convergence of the screening matrix elements with large $\mathbf{G}_1$ or $\mathbf{G}_2$ is expected to be governed by the inclusion of high energy (free-electron like) KS states in the sum over empty states.

In 62\% of the systems the $\Gamma - \Gamma$ gap converges towards a larger gap in both parameters as is the case for BN (Fig. ~\ref{plot-2dconv}). In the remaining systems we observe opposite convergence directions for $E_\mathrm{c}$ and $N_\mathrm{b}$, where $E_\mathrm{c}$ is slightly more often converging towards a larger gap. In about 23\% of the systems some form non-monotonous behaviour is observed. In the cases where the convergence in $N_\mathrm{n}$ is non-monotonous this happens at low number of bands and is rather sharp, as is seen for Si in Fig.~\ref{plot-2dconv}. The exact nature of this non-monotonous behaviour in $N_\mathrm{b}$ seems to depend on particular features in the band structure that are hard to predict a priori. In the cases where it occurs in $E_\mathrm{c}$ it occurs at high $E_\mathrm{c}$ values and is usually not very strong, creating just a slight maximum value for the gap. There are even cases (e.g. WS$_2$) where the gap converges with respect to $E_\mathrm{c}$, upwards for low values of $N_\mathrm{b}$ and downwards for high values of $N_\mathrm{b}$.

\begin{figure}[!ht]
\centering
\includegraphics[width=0.90\columnwidth]{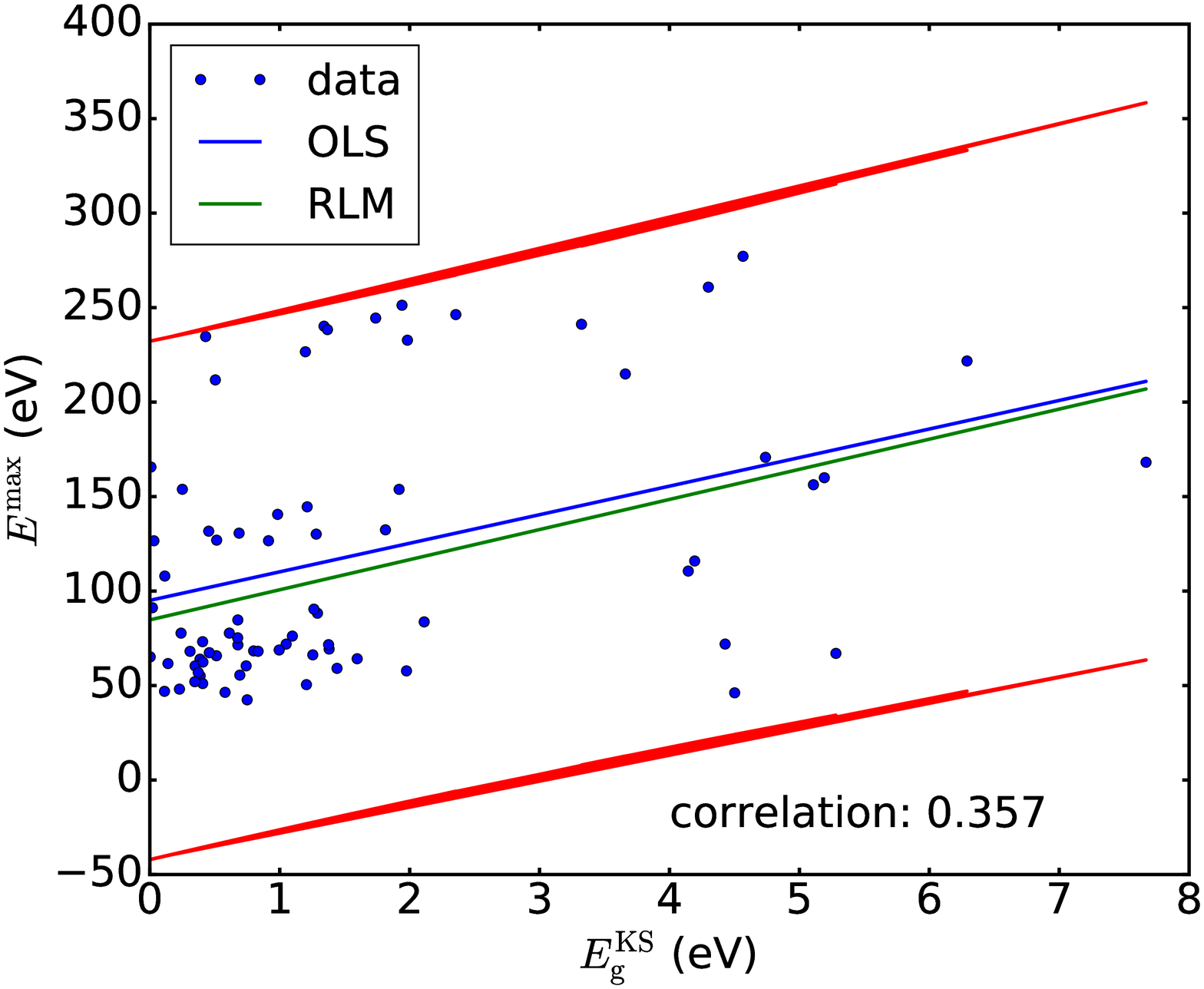}
\caption{Comparison of the energy of the highest band included in the construction of $\Sigma$ needed to reach convergence and the KS-gap.}\label{plot-emvsks} 
\end{figure}

The correlation between the KS-gap and the energy of the highest band, see Fig.~\ref{plot-emvsks}, is the second significant correlation. But, with a correlation coefficient of only 0.35, it clearly looks less promising. The large fraction of outliers and large spread make it difficult to extract any heuristic rule. 

\section{Conclusions}
 
We have presented a methodology to perform $GW$ calculations in an automatic manner requiring as little as possible human intervention. The most important aspects are the coupled 2D convergence study on the number of empty states $N_\mathrm{b}$ and the energy cutoff of the screening and self-energy $E_\mathrm{c}$ and the decoupling of this convergence study and the density of the k-point mesh. We apply this methodology to $\sim$80 solids establishing the accuracy and convergence properties of $G_0W^\mathrm{GN}0$@PBE: single shot perturbative $G_0W_0$ using the KS-PBE\cite{tmf8} as a starting point within the Godby-Needs plasmon-pole model.\cite{godby:prb88}

We confirm that the convergence behavior of $N_\mathrm{b}$ and $E_\mathrm{c}$ are connected. For a large part of the systems in our study we observe that studying the convergence of one at low value of the other leads to under-converged parameters. We observe a small positive correlation between $E_\mathrm{c}$ and $N_\mathrm{b}$. It indicates that a convergence study could be done on a single parameter keeping the relation between the two fixed. This, however, does not hold in general. A single-parameter convergence study may hence lead to over-converged parameters leading to unnecessarily high computational costs.

The correlation between the $G_0W_0^\mathrm{GN}$@PBE error (the difference between the $G_0W_0^\mathrm{GN}0$@PBE and experimental gap) and the experimental gap is clear if the compounds are separated based on whether or not transition metals are present in the compound. For the transition metal compounds we find the relation \TMRLMfunc . For the non-transition metal compounds we observe a smaller error and a less strong dependency  \nonTMRLMfunc .  

The correlation between the QP and KS gaps is very strong,  \qpvsksOLSfunc , and contains almost no outliers. Moreover, it does not show a separation between the transition metal and non transition metal containing compounds. The error of $G_0W_0^{\mathrm{GN}}$@PBE with respect to experiment (MAE of 0.64 and 0.38~eV, TM and nonTM compounds respectively) is actually larger than the error of approximating the QP gap via its relation with the KS gap (MAE of 0.37~eV and 0.29~eV, TM and nonTM compounds respectively).

Finally, we build a linear model to predict the experimental gaps from the calculated KS gaps and compare the predictive power of this model to that of $G_0W_0^{\mathrm{GN}}$@PBE. We find that both the mean absolute error and the standard deviation of the $GW$ results are smaller than those of the linear model. 

\begin{acknowledgments}
The present research benefited from computational resources made available on the Tier-1
supercomputer of the F\'{e}d\'{e}ration Wallonie-Bruxelles, infrastructure funded by the Walloon Region
under the grant agreement n°1117545. Financial support for MG was provided from FRS-FNRS through the PDR Grant T.0238.13 (AIXPHO). The work was supported by the Communaut\'{e} fran\c{c}aise de Belgique through the BATTAB project (ARC 14/19-057).
\end{acknowledgments}

\bibliography{bibfile}

\end{document}

%% file: newcommands.tex
\newcommand{\std}{0.64}

\newcommand{\gwkscor}{0.99}
\newcommand{\ecutepsmaxencor}{0.53}
\newcommand{\mean}{-0.21}

\newcommand{\TMRLMfunc}{$E_\mathrm{g}^\mathrm{GW}$-$E_\mathrm{g}^\mathrm{EXP}$~=~0.29~-~0.28$E_\mathrm{g}^\mathrm{EXP}$}

\newcommand{\nonTMfunc}{$E_\mathrm{g}^\mathrm{GW}$-$E_\mathrm{g}^\mathrm{EXP}$~=~0.09~-~0.07$E_\mathrm{g}^\mathrm{EXP}$ (R$^2$ = 0.16)}

\newcommand{\TMfunc}{$E_\mathrm{g}^\mathrm{GW}$-$E_\mathrm{g}^\mathrm{EXP}$~=~0.42~-~0.37$E_\mathrm{g}^\mathrm{EXP}$ (R$^2$ = 0.14)}

\newcommand{\qpvsksOLSfunc}{$E_\mathrm{g}^\mathrm{GW}$~=~0.51~+~1.32$E_\mathrm{g}^\mathrm{KS}$ (R$^2$ = 0.97)}

\newcommand{\qpvsksRLMfunc}{$E_\mathrm{g}^\mathrm{GW}$~=~0.53~+~1.31$E_\mathrm{g}^\mathrm{KS}$}
\newcommand{\nonTMRLMfunc}{$E_\mathrm{g}^\mathrm{GW}$-$E_\mathrm{g}^\mathrm{EXP}$~=~0.13~-~0.07$E_\mathrm{g}^\mathrm{EXP}$}
\newcommand{\systems}{Ca$_2$Sn, Ca$_2$Pb, As$_2$Os, RuS$_2$, CdTe, CuCl, CuBr, CuI, AgI, CsAu, CaO, SrO, C, BAs, AlAs, GaP, InP, MgO, K$_3$Sb, BP, CdSe, Mg$_2$Sn, InSe, GaSe, PbS, PbSe, SnSe$_2$, MoS$_2$, AgCl, AgBr, RbAu, Bi$_2$Te$_3$, TiO$_2$, NaCl, CsCl, CsBr, ZnS, ZnSe, ZnTe, HgS, Si, Ge, SiC, BN, BN, AlN, AlSb, GaN, GaAs, GaSb, BeO, AgI, Ga$_2$S$_3$, Cs$_3$Sb, Ag$_2$O, CdP$_4$, TePb, SnO$_2$, PbF$_2$, NiP$_2$, WS$_2$, BeTe, HgO, TlSe, Na$_3$Sb, Rb$_3$Sb, GeSe, SnSe, K$_3$Sb, GeS, PbO, ZrSe$_3$, TlS, Al$_2$O$_3$, PbO, PtS, KCl, V$_2$O$_5$}

%% file: table_full_describe.tex
\begin{tabular}{lrrrrrr}
\toprule
{} &  $E_\mathrm{g}^\mathrm{EXP}$ &  $E_\mathrm{g}^\mathrm{KS}$ &  $E_\mathrm{g}^\mathrm{GW}$ &  $E_\mathrm{g}^\mathrm{GW}$-$E_\mathrm{g}^\mathrm{EXP}$ &  $Z^\mathrm{max}$ &  $Z^\mathrm{min}$ \\
\midrule
count &                        75.00 &                       75.00 &                       75.00 &                                              75.00 &             75.00 &             75.00 \\
mean  &                         2.81 &                        1.63 &                        2.62 &                                              -0.18 &             46.40 &             21.20 \\
std   &                         2.38 &                        1.69 &                        2.26 &                                               0.63 &             23.01 &             13.93 \\
min   &                         0.13 &                        0.02 &                        0.44 &                                              -2.68 &              6.00 &              4.00 \\
25\%  &                         1.15 &                        0.48 &                        1.09 &                                              -0.35 &             31.00 &              8.00 \\
50\%  &                         2.10 &                        1.00 &                        1.93 &                                              -0.09 &             49.00 &             16.00 \\
75\%  &                         3.17 &                        1.93 &                        2.94 &                                               0.15 &             55.00 &             32.50 \\
max   &                        10.59 &                        7.67 &                       10.10 &                                               1.30 &             83.00 &             55.00 \\
\bottomrule
\end{tabular}

%% file: table_outliers.tex
\begin{tabular}{lrrrrrrr}
\toprule
 system &  type &  $E_\mathrm{g}^\mathrm{EXP}$ &  $E_\mathrm{g}^\mathrm{KS}$ &  $E_\mathrm{g}^\mathrm{GW}$ &  $E_\mathrm{g}^\mathrm{GW}$-$E_\mathrm{g}^\mathrm{EXP}$ &  $Z^\mathrm{max}$ &  $Z^\mathrm{min}$ \\
\midrule
   CuCl &     2 &                         3.30 &                        0.50 &                        0.62 &                                              -2.68 &                29 &                17 \\
   CuBr &     2 &                         2.91 &                        0.43 &                        0.64 &                                              -2.27 &                35 &                29 \\
    GeS &     1 &                         1.65 &                        0.00 &                        0.00 &                                              -1.65 &                32 &                16 \\
   SnO$_2$ &     1 &                         3.73 &                        1.21 &                        2.34 &                                              -1.39 &                50 &                 8 \\
   RuS$_2$ &     2 &                         1.80 &                        0.69 &                        0.47 &                                              -1.33 &                44 &                16 \\
    CuI &     2 &                         2.95 &                        1.10 &                        1.79 &                                              -1.16 &                53 &                29 \\
   RbAu &     2 &                         2.50 &                        0.25 &                        1.37 &                                              -1.13 &                79 &                37 \\
    GaN &     1 &                         3.44 &                        1.94 &                        2.31 &                                              -1.13 &                31 &                 7 \\
    CaO &     1 &                         6.93 &                        3.66 &                        5.86 &                                              -1.07 &                20 &                 8 \\
   NaCl &     1 &                         8.97 &                        5.11 &                        8.01 &                                              -0.96 &                17 &                11 \\
   NiP$_2$ &     2 &                         0.73 &                        0.24 &                       -0.21 &                                              -0.94 &                28 &                15 \\
 Bi$_2$Te$_3$ &     1 &                         0.13 &                        0.35 &                        0.58 &                                               0.45 &                83 &                52 \\
   TlSe &     1 &                         0.73 &                        0.41 &                        1.21 &                                               0.48 &                81 &                34 \\
    PtS &     2 &                         0.80 &                        0.45 &                        1.30 &                                               0.50 &                78 &                16 \\
    BAs &     1 &                         1.46 &                        1.20 &                        2.11 &                                               0.65 &                33 &                 5 \\
   BeTe &     1 &                         2.80 &                        1.98 &                        3.45 &                                               0.65 &                52 &                 4 \\
  SnSe$_2$ &     1 &                         0.97 &                        0.68 &                        1.68 &                                               0.71 &                50 &                34 \\
   TePb &     1 &                         0.31 &                        0.74 &                        1.05 &                                               0.74 &                82 &                52 \\
   V$_2$O$_5$ &     2 &                         2.20 &                        1.81 &                        3.50 &                                               1.30 &                23 &                 8 \\
\bottomrule
\end{tabular}

%% file: table_descibe_convpar.tex
\begin{tabular}{lrrrr}
\toprule
{} &  $E_\mathrm{c}$ &  $E_\mathrm{c}^\mathrm{interpol}$ &  $N_\mathrm{b}$ &  $E^\mathrm{max}$ \\
\midrule
count &           78.00 &                             78.00 &           78.00 &             78.00 \\
mean  &          244.28 &                            150.26 &          282.82 &            117.79 \\
std   &           60.45 &                             78.29 &          278.36 &             72.43 \\
min   &           93.11 &                             93.87 &           35.00 &             33.09 \\
25\%  &          212.05 &                            105.33 &          105.00 &             64.43 \\
50\%  &          216.66 &                            108.38 &          215.00 &             80.71 \\
75\%  &          315.46 &                            178.46 &          342.50 &            155.65 \\
max   &          432.87 &                            434.53 &        1,505.00 &            338.67 \\
\bottomrule
\end{tabular}